# Towards a direct measurement of vacuum magnetic birefringence: PVLAS achievements

F. Della Valle³, G. Di Domenico¹, U. Gastaldi², E. Milotti³, R. Pengo², G. Ruoso², G. Zavattini¹

¹ *INFN, Sezione di Ferrara and Dipartimento di Fisica, Università di Ferrara, Via Saragat 1, Blocco C, 44100 Ferrara, Italy*
² *INFN, Laboratori Nazionali di Legnaro, viale dell'Università 2, 35020 Legnaro, Italy*
³ *INFN, Sezione di Trieste and Dipartimento di Fisica, Università di Trieste, Via Valerio 2, 34127 Trieste, Italy*

**Abstract:** Nonlinear effects in vacuum have been predicted but never observed yet directly. The PVLAS collaboration has long been working on an apparatus aimed at detecting such effects by measuring vacuum magnetic birefringence. Unfortunately the sensitivity has been affected by unaccounted noise and systematics since the beginning. A new small prototype ellipsometer has been designed and characterized at the Department of Physics of the University of Ferrara, Italy entirely mounted on a single seismically isolated optical bench. With a finesse $F = 414000$ and a cavity length $L = 0.5$ m we have reached the predicted sensitivity of $\psi = 2 \cdot 10^{-8}$ $1/\sqrt{\text{Hz}}$ given the laser power at the output of the ellipsometer of $P = 24$ mW. This record result demonstrates the feasibility of reaching such sensitivities and opens the way to designing a dedicated apparatus for a first detection of vacuum magnetic birefringence.

**Introduction**

Vacuum magnetic birefringence and elastic light-light scattering have been predicted many years ago. Both effects are associated to the electron-positron vacuum fluctuations and can be calculated in the framework of the Euler-Heisenberg-Weisskopf effective Lagrangian correction $L_{EHW}$ (S.I. units):

$$L_{EHW} = \frac{1}{2\mu_0}\left(\frac{\vec{E}^2}{c^2} - \vec{B}^2\right) + \frac{A_e}{\mu_0}\left[\left(\frac{\vec{E}^2}{c^2} - \vec{B}^2\right)^2 + 7\left(\vec{E} \cdot \vec{B}\right)^2\right] \tag{1}$$

where the parameter $A_e$ describing the nonlinear behaviour is

$$A_e = \frac{2}{45\pi}\frac{\alpha^2 \lambdabar_e^3}{m_e c^2} = 1.32 \cdot 10^{-24} \text{ T}^{-2} \tag{2}$$

As a consequence, it can be shown that a region filled with a uniform magnetic field becomes birefringent [1,2] and that the refractive index difference for the linear polarizations parallel and perpendicular to the field is,

$$\Delta n = 3A_e B_0^2 \tag{3}$$

which for a 2.3 T magnetic field results in $\Delta n = 2.1 \cdot 10^{-23}$. Similarly one can also calculate the total elastic light-light scattering cross section for unpolarized light which results proportional to $A_e^2$:

$$\sigma_{\gamma\gamma}^{(QED)}(\hbar\omega) = \frac{973\mu_0^2}{20\pi}\frac{\hbar^2 \omega^6}{c^4}A_e^2 \tag{4}$$

For $\lambda = 1064$ nm this cross section is $\sigma_{\gamma\gamma} = 1.8 \cdot 10^{-69}$ m².

These values show the extreme difficulty of measuring such effects and explains why they still have to be measured.

At present the best limits obtained experimentally are from the PVLAS collaboration and can be summarized in the parameter $A_e$ measured with a field strength of $B_0 = 2.3$ T:

$$A_e^{(PVLAS)} < 6.6 \cdot 10^{-21} \text{ T}^{-2} \text{ @ 1064 nm}$$
$$A_e^{(PVLAS)} < 6.3 \cdot 10^{-21} \text{ T}^{-2} \text{ @ 532 nm}$$
(5)

The average measured sensitivity to ellipticity of the apparatus present at the Laboratori Nazionali di Legnaro, Padova, Italy during long runs was about $\psi^{(PVLAS)} \approx 1 \cdot 10^{-6}$ 1/√Hz, a factor between 50 and 100 worse than the predicted sensitivity given the apparatus in ref [3]. This is shown in figure 1 for both the $\lambda = 1064$ nm and $\lambda = 532$ nm where the sensitivity is plotted as a function of the modulation amplitude of the modulator (see section **Apparatus and Method** below).

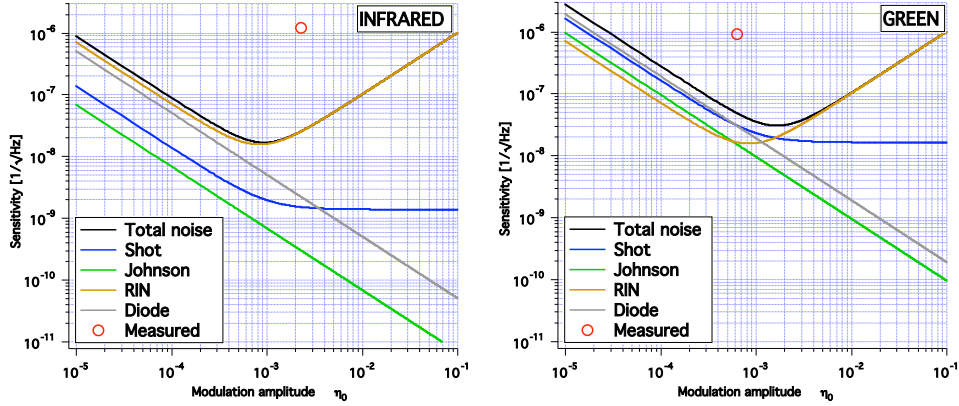

Figure 1: Calculated and measured sensitivity of the PVLAS apparatus for the $\lambda = 1064$ nm (left) and the $\lambda = 532$ nm (right) cases.

The Legnaro ellipsometer is a rather bulky structure [3] because it was designed to host a powerful superconducting magnet in a LHe cooled cryostat. Due to the low duty cycle of the Legnaro ellipsometer it would be difficult to significantly improve the present limits so we undertook a study to understand the underlying noise affecting the ellipsometer. The PVLAS collaboration therefore designed a new small 50 cm long tabletop ellipsometer based on the same measurement principle as the apparatus at the Laboratori Nazionali di Legnaro, Padova, Italy.

**Apparatus and Method**

The working principle of these ellipsometers is shown in figure 2. A polarizer **P** defines the linear polarization of the incoming laser beam. Two mirrors **M1** and **M2** define a Fabry-Perot optical cavity with very high finesse F. Part of the region between the mirrors is the sensitive region for ellipsometric measurements and the ellipticity $\psi$ acquired in a single pass is multiplied by a factor $2F/\pi$: $\Psi = \psi \, (2F/\pi)$. After mirror **M2** a photoelastic modulator **MOD** adds a known ellipticity $\eta$ to the already acquired total ellipticity $\Psi$. After the modulator an analyzer **A**, rotated at 90° with respect to **P**, selects the electric field component perpendicular to the input polarization. The transmitted light is then detected by a low noise photodiode.

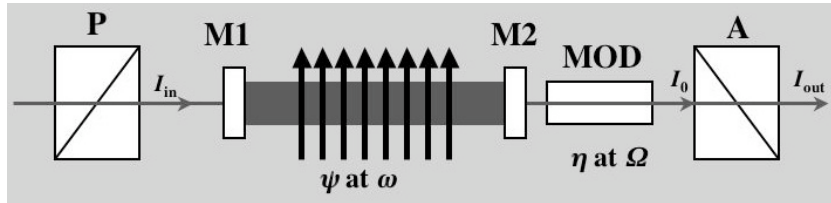

Figure 2: Scheme of the ellipsometer principle studied by the PVLAS collaboration.

Both $\Psi$ and $\eta$ can be varied in time. Experimentally, spurious noise sources are also present: slowly varying ellipticities $\alpha(t)$ are present in all optical elements; polarizers are non ideal and have extinction ratio $\sigma^2$. The total measured intensity can therefore be written as

$$I_{out} = I_0 \left\{ \sigma^2 + \eta(t)^2 + 2\eta(t)\alpha(t) + 2\eta(t)\Psi(t) + ... \right\} \tag{6}$$

The introduction of the modulator linearizes the effect $\Psi(t)$ which would otherwise be quadratic. Since the ellipticity $\Psi(t)$ varies sinusoidally with angular frequency $\omega$ and the modulation with angular frequency $\Omega$ then the main Fourier components in the photodiode current are those reported in table 1.

Table 1: Main Fourier frequency components in the intensity signal at the output of the ellipsometer.

| Angular Frequency | Fourier component | Intensity/$I_0$ |
|---|---|---|
| dc | $I_{DC}$ | $\sigma^2 + \alpha_{DC}^2 + \eta_0^2/2$ |
| $\omega$ | $I_\omega$ | $2\alpha_{DC}\eta_0$ |
| $\Omega \pm \omega$ | $I_{\Omega \pm \omega}$ | $\Psi\eta_0$ |
| $2\Omega$ | $I_{2\Omega}$ | $\eta_0^2/2$ |

If the signal $\Psi$ is above the noise it can be extracted from the Fourier components as the average value of the sideband signals $I_{\Omega\pm\omega}$:

$$\Psi = \frac{1}{2}\left( \frac{I_{\Omega+\omega}}{\sqrt{2I_0 I_{2\Omega}}} + \frac{I_{\Omega-\omega}}{\sqrt{2I_0 I_{2\Omega}}} \right) \tag{7}$$

Several experimental efforts – both present and past - are based on this principle. The first was the BFRT collaboration (Brookhaven, Fermilab, Rochester, Trieste). In their setup a multipass non resonant cavity was used. Current experiments all use Fabry-Perot cavities since the number of equivalent passes can be as high as 260000 (reported here). One aspect which is common to all these experiments is that the theoretical sensitivity could not be reached with the optical path multipliers inserted. In the table 2, below, we report the sensitivities of the different efforts with the corresponding number of passes.

Table 2: Rotation and ellipticity sensitivities of various experimental efforts using ellipsometers with optical path multipliers

| Esperiment | number of passes | ellipticity/rotation sensitivity [$1/\sqrt{Hz}$] | sensitivity without cavity [$1/\sqrt{Hz}$] | $\Delta n$ [$1/\sqrt{Hz}$] |
|---|---|---|---|---|
| BFRT [5] | 254 | $7 \cdot 10^{-8}$ @ 30 mHz (rotation) | $7 \cdot 10^{-9}$ @ 30 mHz (rotation) | $1.6 \cdot 10^{-17}$ |
| Q&A [6] | 10800 | $10^{-6}$ @ 10 Hz (ellipticity) | - | $9 \cdot 10^{-18}$ |
| PVLAS LNL [3] | 45000 | $10^{-6}$ @ 0.6 Hz (ellipticity) | - | $1.1 \cdot 10^{-18}$ |
| J. Hall [7] | 26000 | $3 \cdot 10^{-5}$ @ 2 mHz (ellipticity) | - | $8 \cdot 10^{-16}$ |
| PVLAS Ferrara | 260000 | $2 \cdot 10^{-8}$ @ 5 Hz (ellipticity) | $5 \cdot 10^{-9}$ @ > 2 Hz (ellipticity) | $5 \cdot 10^{-20}$ |

Following the schematic layout of figure 2, a new compact 50 cm long benchtop ellipsometer was assembled at the Department of Physics of the University of Ferrara, Italy, to study the noise sources present in ellipsometers based on Fabry-Perot cavities. The vibrational nature of the noise in the Legnaro ellipsometer has been suspected for a long time. The new setup, therefore, was designed to be used both with and without seismic isolation; the basic seismic isolation uses pneumatic legs, which act as a low-pass filter with a cutoff frequency of about 5 Hz. At first, for a less critical setup, we used a cavity with finesse about $F = 3000$.

After achieving the best possible result, we mounted higher finesse mirrors and reached a value of $F = 414000$, as can be seen in figure 3. We also achieved an input coupling of 75% with a total transmission of 25%. In the 50 cm long benchtop ellipsometer the optical elements of the ellipsometer, including the polarizer, cavity mirrors, modulator and analyzer, were mounted on a single breadboard. The distance between the mirrors of the Fabry-Perot cavity was about $L = 50$ cm. The entire breadboard was then inserted in a vacuum chamber. The polarizer **P**, entrance mirror **M1**, and analyzer **A** were mounted on motorized rotating mounts for polarization alignment and to achieve the best possible extinction ratio with the cavity inserted. We used the standard Pound-Drever-Hall method to lock the laser to the cavity but with the laser itself being used as the optical phase modulator [4]. The laser is a 200 mW Innolight Nd:YAG 1064 nm laser.

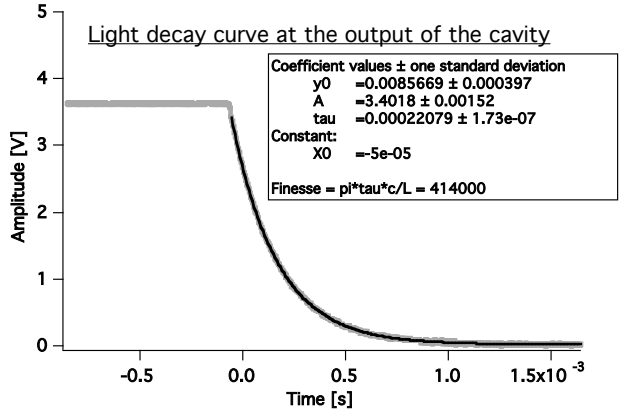

Figure 3: Light decay curve at the output of the high finesse cavity (grey) with superimposed an exponential fit (black curve). Given the cavity length of about $L = 50$ cm the resulting finesse was $F = 414000$.

Using the experimental parameters of the 50 cm long ellipsometer the achievable sensitivity as a function of the modulation amplitude can be determined [3] for both the high and low finesse configurations. These are reported in figure 4. In principle, in the high finesse configuration a sensitivity of about $3 \cdot 10^{-9}$ $1/\sqrt{Hz}$ should be achievable. As can be seen in figure 4, right, due to the low pass filter nature of the high finesse cavity, residual intensity noise at the modulator carrier frequency of 50 kHz is greatly suppressed.

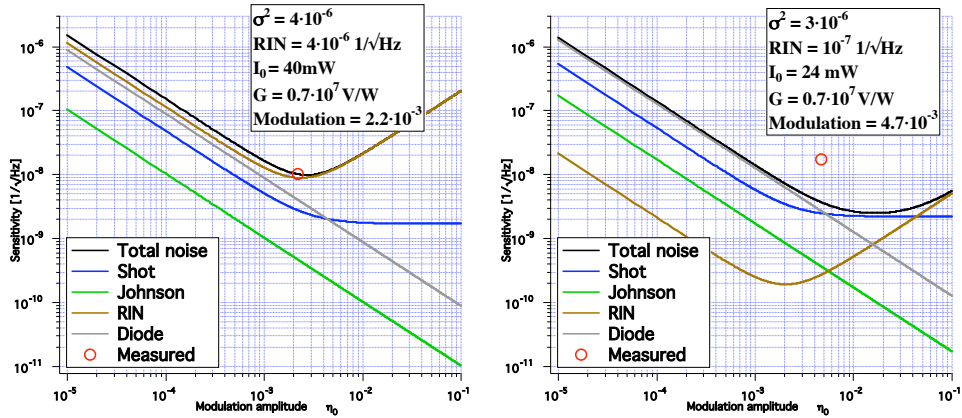

Figure 4: Sensitivity plots for the low finesse (left) and high finesse (right) configurations as a function of the modulation amplitude. Shot-noise, Johnson noise, residual intensity noise (RIN), and photodiode noise contributions are shown separately. In black is shown the total noise. The red circle indicates the measured sensitivity.

**Results**

In figure 5 we report the acceleration spectral densities for the optical bench used. The top curve corresponds to the vertical acceleration when the seismic isolation is inactive. The two lower curves correspond to the vertical and horizontal acceleration of the bench with the isolation active. In the frequency range 5 Hz – 50 Hz there is a suppression factor between 10 and 50. As far as the present work is concerned the region of interest is between 5 Hz and 20 Hz.

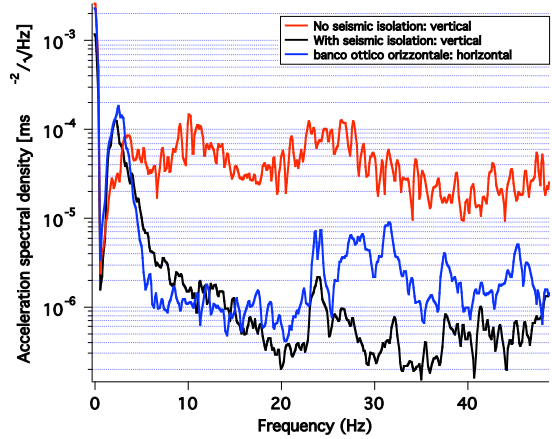

Figure 5: measured acceleration spectral densities. The top red curve corresponds to the configuration with the seismic isolation turned OFF whereas the blue and black curves correspond respectively to the vertical and horizontal accelerations with the seismic isolation turned ON.

Figure 6 shows the ellipticity spectral density for the low finesse configuration both with and without seismic isolation. There is a clear improvement of about a factor 10 in the frequency range of interest when the seismic isolation is turned ON. We also remark that when the seismic isolation is turned ON most spectral structures disappear and the noise is flat in the frequency range of interest.

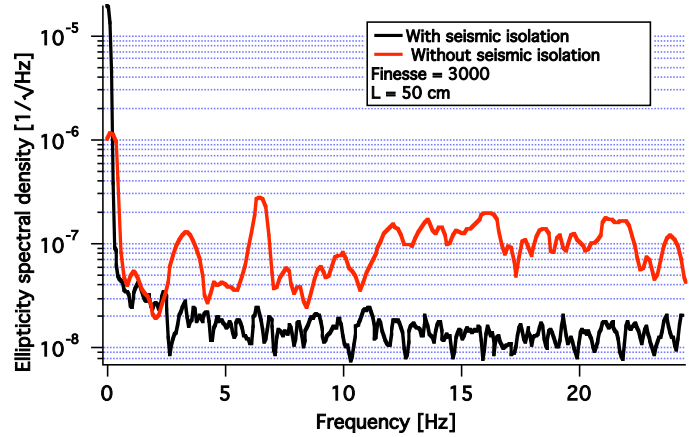

Figure 6: Single-sided (only frequencies above the carrier frequency) measured ellipticity sensitivity plots as a function of frequency from the modulator carrier frequency. The upper red curve corresponds to the configuration with seismic isolation turned OFF whereas the black curve corresponds to the configuration with the seismic isolation turned ON.

The average noise of the single sided spectrum (only frequencies above $\Omega_{Mod}$) between 5 Hz and 20 Hz is $1.4 \cdot 10^{-8}$ $1/\sqrt{Hz}$ and averaging with the other half of the spectrum one obtains a sensitivity of

$$\Psi_{3000} = 1 \cdot 10^{-8} \ \frac{1}{\sqrt{\text{Hz}}} \tag{8}$$

Comparing this measured value with the expected value superimposed in figure 4 as a red circle it is clear that we have reached the expected theoretical value of the sensitivity. In this case sensitivity is limited by the residual intensity noise at the modulation frequency.

In figure 7 we show the spectral density of ellipticity noise for the high finesse configuration with finesse $F = 414000$. Again the noise spectrum is flat and the noise around the carrier frequency is $2.6 \cdot 10^{-8}$ $1/\sqrt{\text{Hz}}$. By averaging the upper and lower parts of the spectrum we obtain the record sensitivity

$$\Psi_{414000} = 1.8 \cdot 10^{-8} \ \frac{1}{\sqrt{\text{Hz}}} \tag{9}$$

Considering a cavity length of $L = 50$ cm this implies a sensitivity in birefringence of

$$\Delta n_{414000} = 4.6 \cdot 10^{-20} \ \frac{1}{\sqrt{\text{Hz}}} \tag{10}$$

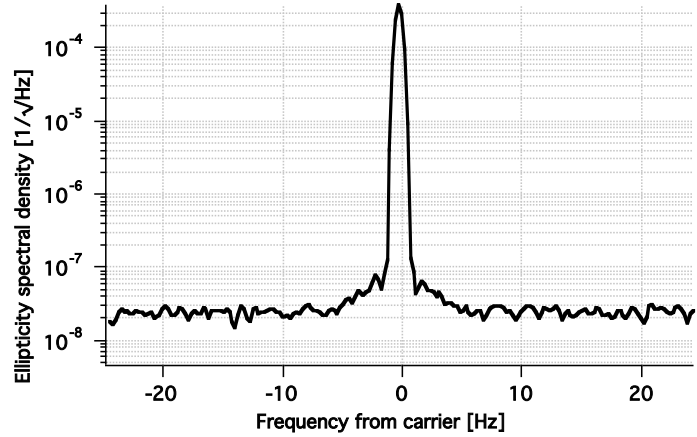

Figure 7: Spectral density of ellipticity noise around the carrier frequency for the high finesse configuration.

**Discussion and developments**

Although the achieved sensitivity is extremely good, it is still about 7 times worse than the shot noise limit, as can be seen in figure 4. Because noise is so flat we can conclude that it cannot have a seismic or mecchanical origin. Indeed these kinds of noises are typically not flat and this is not the case. Furthermore the measured noise is not symmetrical with respect to the carrier frequency and must therefore be a noise floor present at and near the carrier frequency. The origin of this noise is under study.

The 50 cm long ellipsometer with which we have obtained this record sensitivity cannot be used with magnets for magnetic birefringence measurements. Thus a new and slightly longer prototype compatible with the insertion of two dipole magnets has been constructed and is under test. If a similar sensitivity to the one discussed above will be achieved with magnets in place, this will open the way to the construction of a final system for a first vacuum magnetic birefringence measurement. The goal is to be able to measure vacuum magnetic birefringence with a reasonable integration time of $T = 10^6$ s. In figure 8 we show the magnetic field length as a function of sensitivity which is necessary to reach a signal to noise ratio of 1 in $T = 10^6$ s.

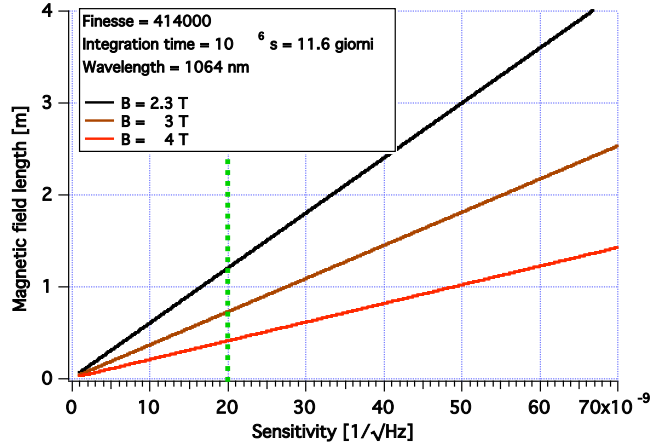

Figure 8: Plots of the necessary magnetic field length as a function of sensitivity to reach a signal to noise ratio of 1 in a time $T = 10^6$ seconds. Three different field strengths are shown: 2.3 T, 3 T and 4 T. The vertical dashed green line indicates the experimentally reached ellipticity sensitivity reported in this paper.

**Conclusion**

In this paper we have reported the sensitivity improvement obtained with a 50 cm long ellipsometer based on a Fabry-Perot optical cavity. The fundamental conclusion for reaching good sensitivities is that the optical bench must be *globally* isolated from seismic noise. With a system having a finesse of $F = 3000$ we were able to reach the theoretical sensitivity based on the experimental parameters. With the system having a finesse $F = 414000$ a record sensitivity of $\Psi_{414000} = 1.8 \cdot 10^{-8}$ $1/\sqrt{Hz}$ was obtained but still greater than the theoretically expected value. Considering a length of $L = 50$ cm this sensitivity implies a birefringence sensitivity of $\Delta n = 4.6 \cdot 10^{-20}$ $1/\sqrt{Hz}$. Hopefully this result will lead to the first measurement of nonlinear QED effects in vacuum.